\documentclass[12pt]{article}

\usepackage[dvips]{graphicx}

\makeatletter
\newtoks\@stequation

\def\subequations{\refstepcounter{equation}%
  \edef\@savedequation{\the\c@equation}%
  \@stequation=\expandafter{\theequation}
  \edef\@savedtheequation{\the\@stequation}
  \edef\oldtheequation{\theequation}%
  \setcounter{equation}{0}%
  \def\theequation{\oldtheequation\alph{equation}}}

\def\endsubequations{%
  \ifnum\c@equation < 2 \@warning{Only \the\c@equation\space subequation
    used in equation \@savedequation}\fi
  \setcounter{equation}{\@savedequation}%
  \@stequation=\expandafter{\@savedtheequation}%
  \edef\theequation{\the\@stequation}%
  \global\@ignoretrue}

\def\eqnarray{\stepcounter{equation}\let\@currentlabel\theequation
\global\@eqnswtrue\m@th
\global\@eqcnt\z@\tabskip\@centering\let\\\@eqncr
$$\halign to\displaywidth\bgroup\@eqnsel\hskip\@centering
     $\displaystyle\tabskip\z@{##}$&\global\@eqcnt\@ne
      \hfil$\;{##}\;$\hfil
     &\global\@eqcnt\tw@ $\displaystyle\tabskip\z@{##}$\hfil
   \tabskip\@centering&\llap{##}\tabskip\z@\cr}
\makeatother

\def\cerenkov{$\check{\rm C}$erenkov~}

\def\eqref#1{eq.~(\ref{#1})}
\def\figref#1{Fig.~\ref{#1}}

\def\PR#1#2#3{Phys. Rev. {\bf #1}, #2 (#3)}
\def\PRL#1#2#3{Phys. Rev. Lett. {\bf #1}, #2 (#3)}
\def\PL#1#2#3{Phys. Lett. {\bf #1}, #2 (#3)}
\def\NL#1#2#3{Nucl. Phys. {\bf #1}, #2 (#3)}
\def\NP#1#2#3{Nucl. Phys. {\bf #1}, #2 (#3)}

\def\PTP#1#2#3{Prog. Theor. Phys. {\bf #1}, #2 (#3)}
\def\EPJ#1#2#3{Eur. Phys. J. {\bf #1}, #2 (#3)}

\def\PRD#1#2#3{Phys. Rev. {\bf D#1},~#2 (#3)}

\def\PLB#1#2#3{Phys. Lett. {\bf B#1} (#2) #3}

\def\etal{{\it et al.}}
\def\ibid{{\it ibid.~}}

\def\simgt{\lower.5ex\hbox{$\; \buildrel > \over \sim \;$}}

\def\sia{$\sin^2 2 \theta_{{\rm atm}}~$}
\def\sis{$\sin^2 2 \theta_{{\rm sol}}~$}
\def\sir{$\sin^2 2 \theta_{{\rm rct}}~$}

\def\dmns{\delta_{_{\rm MNS}}}
\def\cpdelta{$\delta_{_{\rm MNS}}~$}

\def\nn{\nonumber}

\def\bequ{\begin{equation}}
\def\eequ{\end{equation}}
\def\beqn{\begin{eqnarray}}
\def\eeqn{\end{eqnarray}}

\title{Solving the neutrino parameter degeneracy by measuring 
the T2K off-axis beam in Korea}

\author{Kaoru Hagiwara$^{1,2}$, 
Naotoshi Okamura$^3$\thanks{e-mail:~okamura@yukawa.kyoto-u.ac.jp} , 
and 
Ken-ichi Senda$^{1,2}$\thanks{e-mail:~senda@post.kek.jp}
\\ \\
{\it \small $^1$Theory Division, KEK, Tsukuba, 305-0801 Japan }\\
{\it \small $^2$Department of Particle and Nuclear Physics, 
the Graduate University }\\
{\it \small for Advanced Studies (SOKENDAI), Tsukuba, 305-0801 Japan} \\ 
{\it \small $^3$Korea Institute for Advanced Study, }\\
{\it \small 207-43, Cheongnyangni 2dong, Dongdaemun-gu, Seoul, 130-722, Korea.}
}
\date{}

\begin{document}

\maketitle

\vspace{-9.5cm}
\begin{flushright}
KEK-TH-1011
\hspace*{3ex}
KIAS-P05024
\hspace*{3ex}
hep-ph/0504061
\end{flushright}
\vspace{ 9.5cm}
\vspace{-2.0cm}

\begin{abstract}
The T2K neutrino oscillation experiment will start in 2009. 
In this experiment the center of the neutrino beam from J-PARC at Tokai 
village will go through underground beneath Super-Kamiokande, 
reach the sea level east of Korean shore, and an off-axis beam 
at $0.5^{\circ}$ to $1.0^{\circ}$ can be observed in Korea. 
We study physics impacts of putting a 100~kt-level Water \cerenkov 
detector in Korea during the T2K experimental period. 
For a combination of the $3^{\circ}$ off-axis beam at SK with 
baseline length $L = 295$km and the $0.5^{\circ}$ off-axis beam 
in the east coast of Korea at $L = 1000$km, we find that the neutrino 
mass hierarchy (the sign of $m^2_{3} - m^2_1$) can be resolved and 
the CP phase of the MNS unitary matrix can be constrained uniquely 
at 3-$\sigma$ level when \sir $ \simgt 0.06 $.
\end{abstract}
\date{}

\newpage

\par
The results of solar and atmospheric neutrino oscillation experiments 
are consistent with the 3 neutrino model,
so are all the other observations except for the LSND experiments \cite{LSND}.
In this paper we assume the 3 neutrino model,
which has 6 observable parameters in neutrino oscillation experiments.
They are 2 mass squared differences, 3 mixing angles and one CP phase.
The atmospheric neutrino oscillation experiments which measure the 
$\nu_{\mu}$ survival probability determine 
the absolute value of one of the two mass squared-differences
and one mixing angle
\cite{sk-atm04} as
\begin{equation}
\label{atm-data}
1.5\times10^{-3} < 
|m_{3}^2 - m_{1}^2|
< 3.4\times10^{-3} {\mbox{{eV}}}^2
\hspace*{5ex}{\mbox{and}} \hspace*{5ex}
 \sin^22\theta_{\rm atm} > 0.92
\end{equation}
at the 90\% confidence level.
The K2K experiment \cite{k2k04},
which is the first accelerator based long baseline (LBL) neutrino 
oscillation experiment,
confirms the above results.
The solar neutrino experiments,
which measure the $\nu_e$ survival probability in the sun \cite{kayser03},
and the KamLAND experiment which measures the $\bar{\nu}_e$ survival 
probability from reactors \cite{kamland04}, determine the other mass 
squared-difference and another mixing angle as
\begin{equation}
\label{sol-data}
m_2^2 - m_1^2 \equiv  8.2^{+0.6}_{-0.5} \times 10^{-5} {\mbox{{eV}}}^2
\hspace*{5ex}{\mbox{and}} \hspace*{5ex}
 \tan^2\theta_{\rm sol} = 0.40^{+0.09}_{-0.07}\,.
\end{equation}
The CHOOZ reactor experiment \cite{chooz} gives 
the upper bound of the third mixing angle ($\theta_{\rm rct}$) as 
\begin{eqnarray}
&& \sin^22\theta_{\rm rct} < 0.20 \mbox{{~  for  ~}}
|m_{3}^2 - m_{1}^2| = 2.0\times10^{-3}\mbox{{eV}}^2\,,\nn \\
\label{rct-data}
&& \sin^22\theta_{\rm rct} < 0.16 \mbox{{~  for  ~}}
|m_{3}^2 - m_{1}^2| = 2.5\times10^{-3}\mbox{{eV}}^2\,, \\
&& \sin^22\theta_{\rm rct} < 0.14 \mbox{{~  for  ~}}
|m_{3}^2 - m_{1}^2|
 = 3.0\times10^{-3}\mbox{{eV}}^2\,,\nn
\end{eqnarray}
at the 90\% confidence level.
The CP phase ($\dmns$) has not been constrained.
In the future neutrino oscillation experiments,
we should not only measure \sir and \cpdelta, 
but also resolve the parameter degeneracies  
\cite{koike00,minakata01,barger01},
such as the sign of $m_3^2 - m_1^2$.
\par  
There are many next generation LBL neutrino oscillation experiments 
\cite{plans}, which plan to measure the model parameters and to solve 
the parameter degeneracy.
In this paper,
we investigate the possibility of detecting in Korea
the neutrino beam from J-PARC (Japan Proton Accelerator Research Complex)
at Tokai village \cite{j-parc},
that will be available during the period of the T2K (Tokai-to-Kamioka) 
experiment \cite{t2k}.
In the T2K experiment,
the center of the neutrino beam will go through underground beneath 
Super-Kamiokande, 
and reach the 
sea level near the Korean shore.
At the baseline length $L = 295$km away from J-PARC,
the upper side of the beam at $2^\circ$ to $ 3^\circ$ off-axis angle
is observed at Super-Kamiokande (SK),
\begin{figure}
\begin{center}
\includegraphics[angle = 0 , width = 7cm]{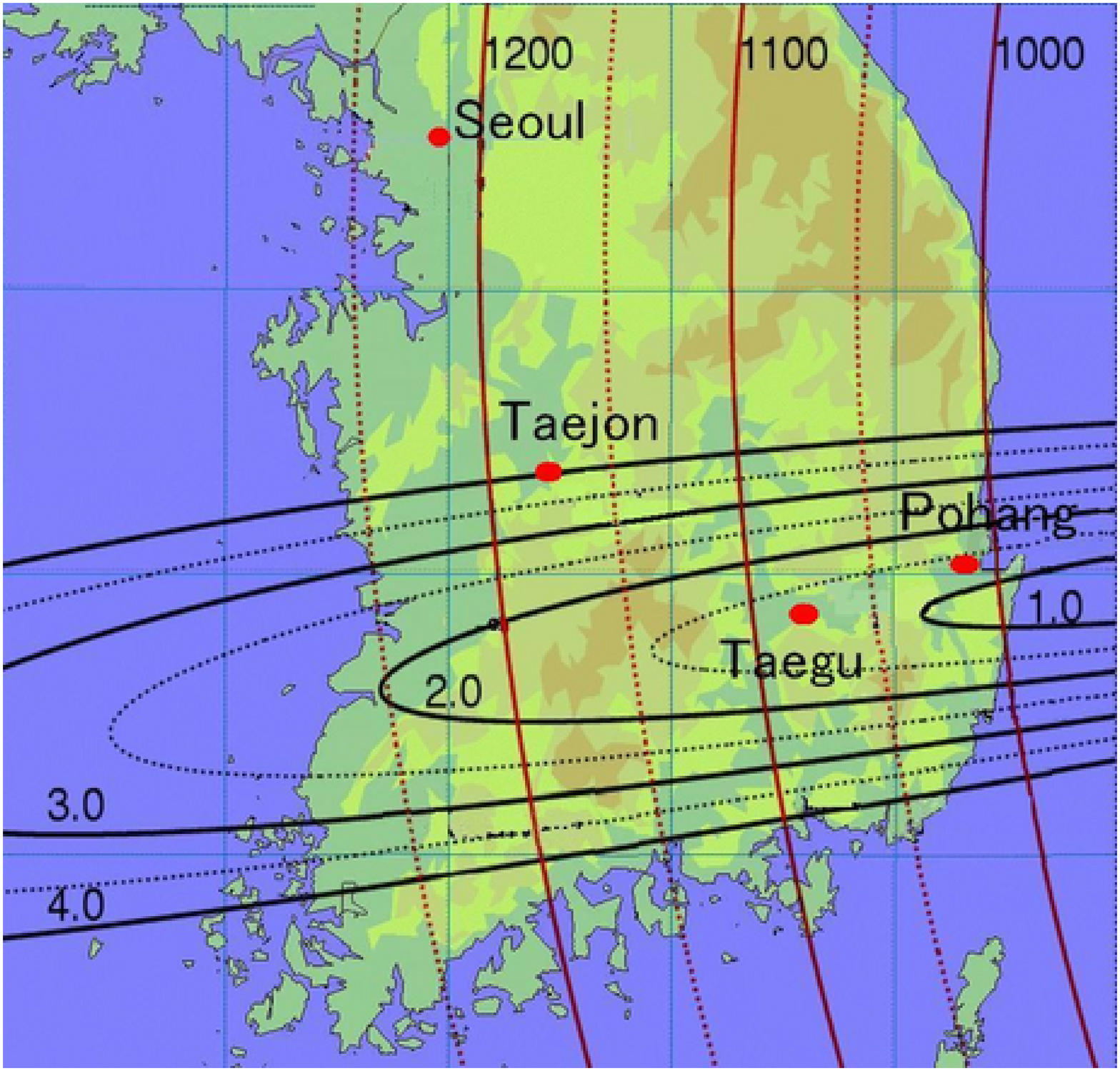}
~~~
\includegraphics[angle = 0 , width = 7cm]{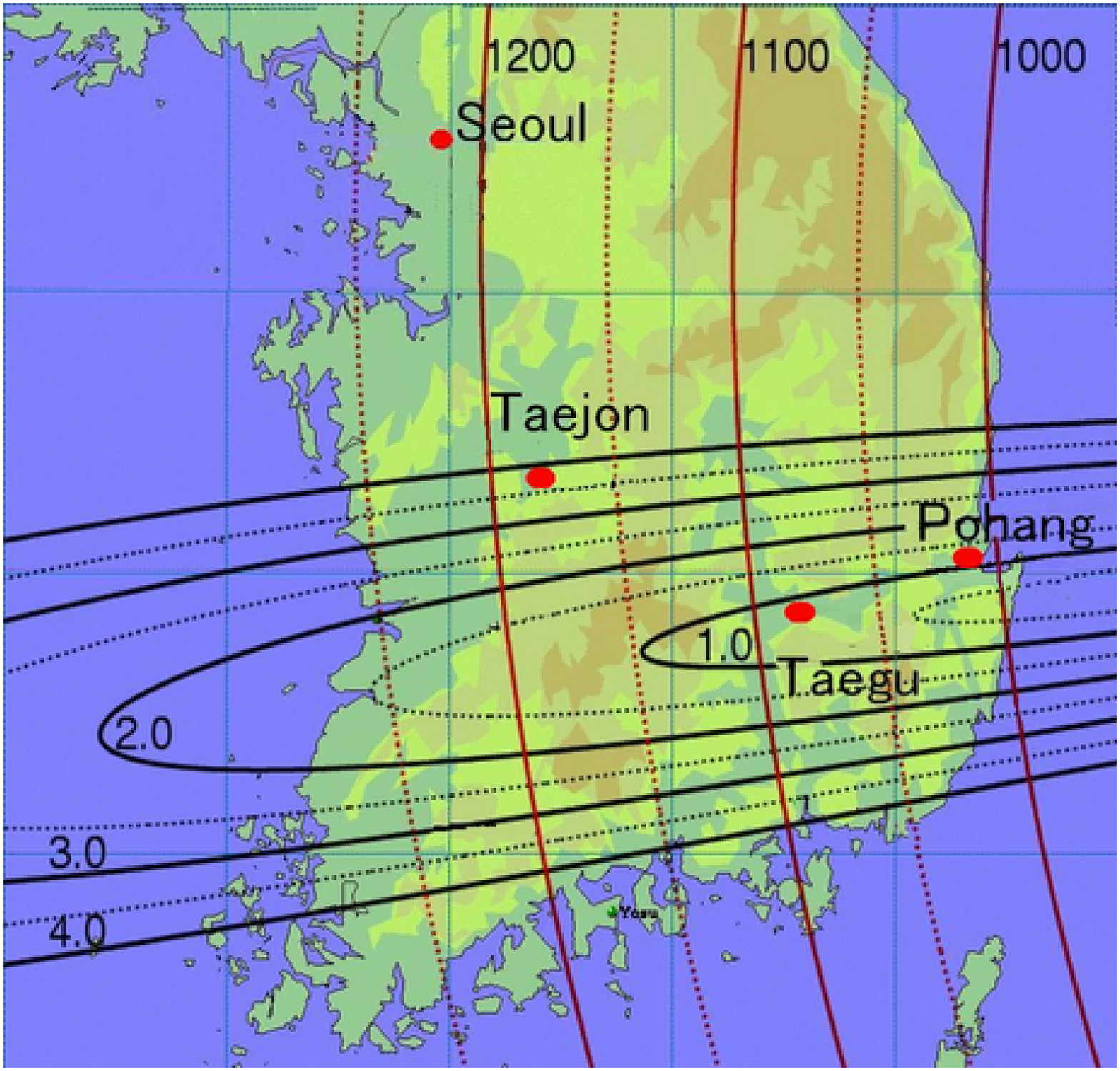}
\end{center} 
\caption{The off-axis angle of the neutrino beam from J-PARC on 
the sea level in Korea, when the beam center is $2.5^{\circ}$(left) 
and $3.0^{\circ}$ (right) off at SK.  
The baseline length for $L = 1000, 1100, 1200$km are shown by 
vertical contours, and the off-axis angles are shown by elliptic 
contours between $0.5^{\circ}$ and $4.0^{\circ}$.
}
\label{fig:map}
\end{figure} 
and the lower side of the same beam at $0.5^\circ$ to $1.0^\circ$
appears in the east coast of Korea \cite{hagiwara04}, 
at about $L = 1000$km; see Fig.~1. 
In order to quantify our investigation,
we study physics impacts of putting a
100kt water \cerenkov detector in Korea during the T2K experiment period, 
which is for 5 years with $10^{21}$ POT (protons on target) per year.
\par
\begin{figure}
\begin{center}
\includegraphics[angle = 0 ,width=7.5cm]{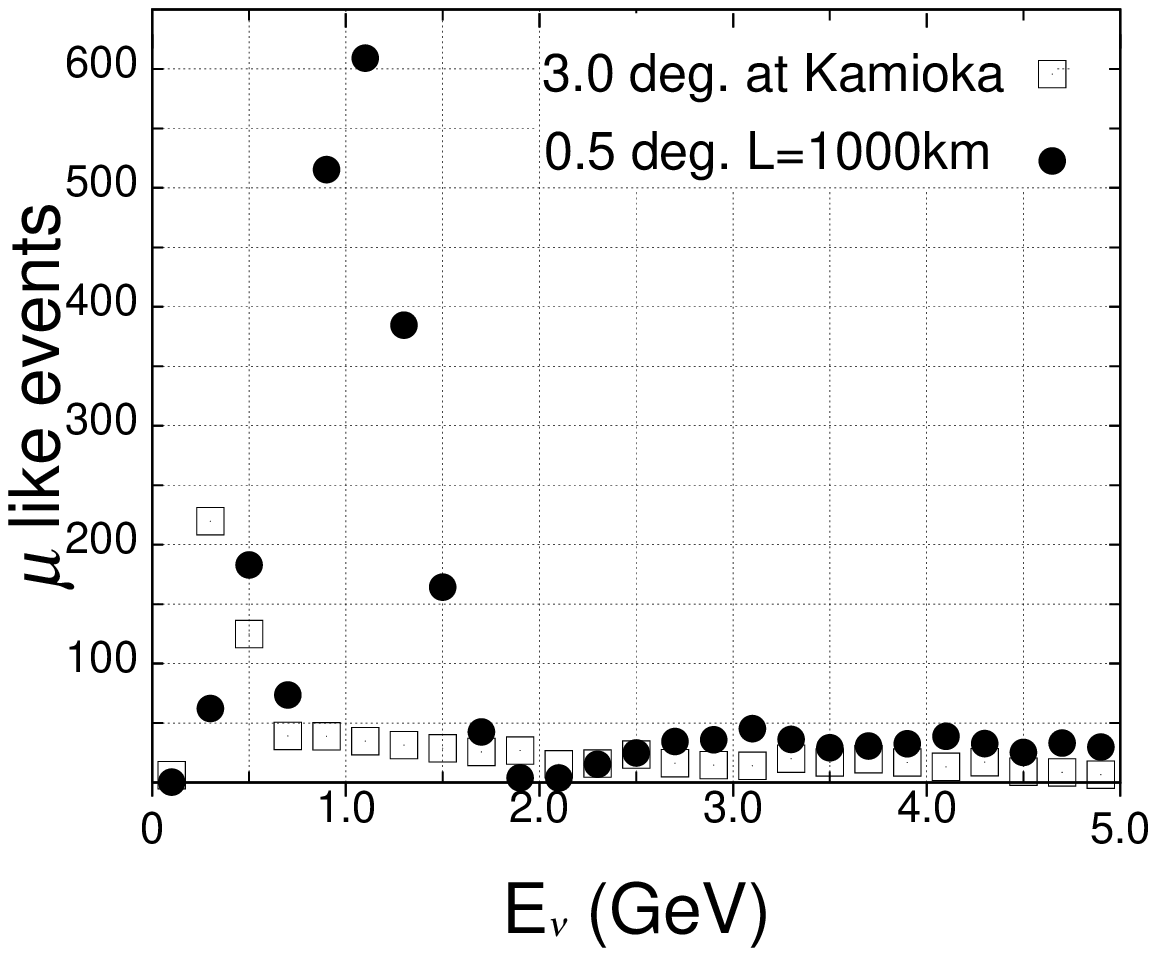}
\includegraphics[angle = 0 ,width=7.5cm]{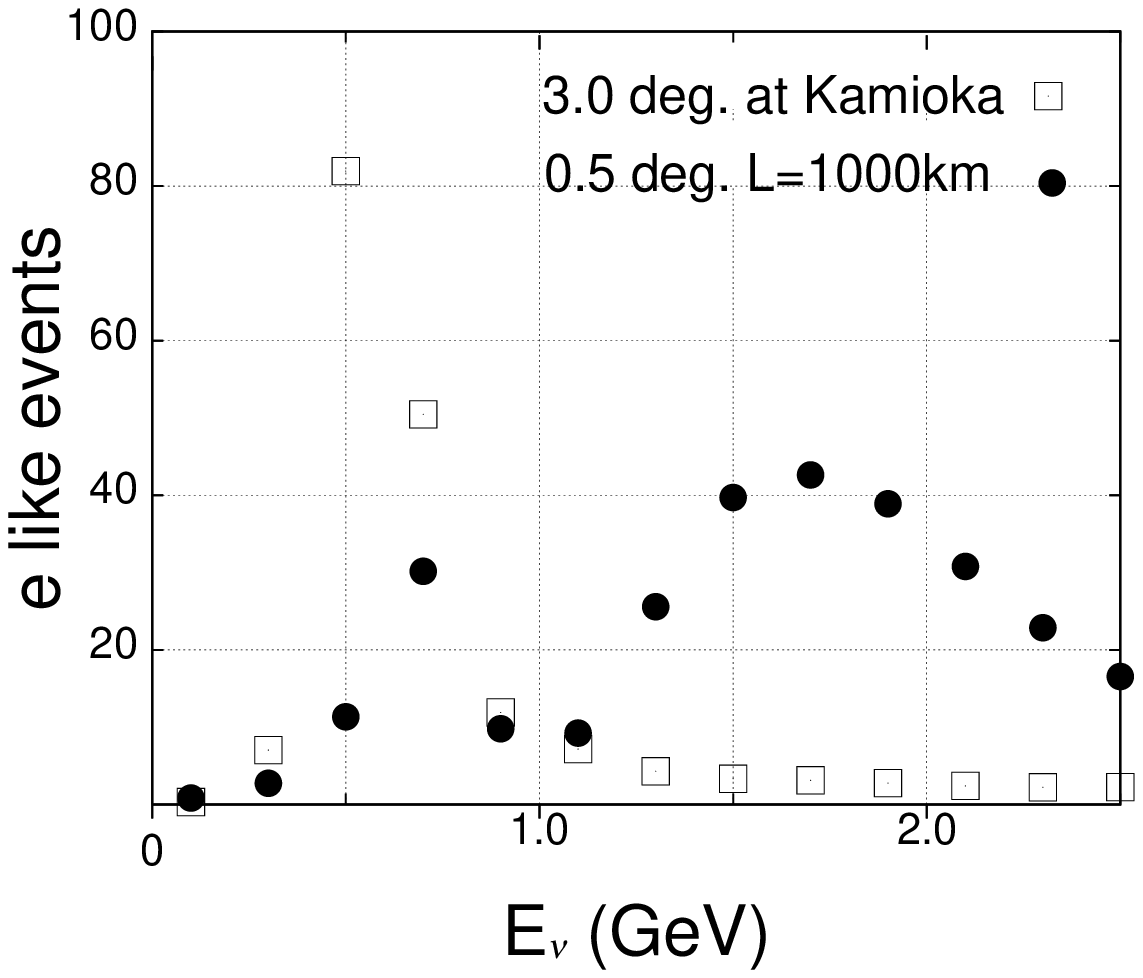}
\end{center}
\caption{The typical numbers of the $\mu$ events (left), and those of the $e$ events (right), for the exposure time of 5 years,
for the $3.0^{\circ}$ OAB at SK (open square), and for the $0.5^{\circ}$ OAB
at $L= 1000~$km with 100kt detector (solid circles). The inputs are $\sin^2 2\theta_{\rm rct}=0.1$ and $\dmns = 0^{\circ}$.
}
\label{fig:event-numbers}
\end{figure}

The merits of measuring the T2K beam in Korea can be summarized as follows:
\begin{enumerate}
  \item Because $0.5^{\circ} - 1.0^{\circ}$ off-axis beam has significantly harder energy spectrum than $2.5^{\circ} - 3.0^{\circ}$ off-axis beam, one can measure the $\nu_{\mu} \rightarrow \nu_e$ transition probability near the oscillation maximum both at Korea and at SK, at the same time.
  \item Because of the matter effect that grows with the baseline length, the difference in the $\nu_{\mu} \rightarrow \nu_e$ transition probability between Korea and SK can reveal the neutrino mass hierarchy pattern \cite{narayan99, super-nova}.
  \item The $\nu_{\mu}$ energy dependence of the oscillation probabilities measured by selecting the quasi-elastic events both at Korea and at SK allows us to constrain both cosine and sine of the CP phase $\dmns$. 
\end{enumerate}
The $\nu_ {\mu}$ survival probability and the $\nu_ {\mu}\rightarrow \nu_e$ transition rate can be expressed as
\begin{eqnarray}
P_{\nu_{\mu} \rightarrow \nu_{\mu}} &=& 1 - \sin^2 2\theta_{\rm atm}\left( 1 + A^{\mu} \right)
\sin^2 \left( \displaystyle\frac{\Delta_{13}}{2} + B^{\mu} \right) \,,
\label{eq:p-numu-numu} \\
P_{\nu_{\mu} \rightarrow \nu_e} &=& 4 \sin^2 \theta_{\rm atm} \sin^2 \theta_{\rm rct}
\left( 1 + A^{e} \right)
\sin^2 \left( \displaystyle\frac{\Delta_{13}}{2} + B^e \right) \,,
\label{eq:p-numu-nue}
\end{eqnarray}
where $\Delta_{ij} =(m^2_j - m^2_i)L/2E$.
Here $A^{\alpha}$ and $B^{\alpha}$ are the corrections to the magnitude and the phase of the oscillation probabilities, respectively,
from the matter effect and the smaller mass-squared difference.
If we keep only those terms linear in $\Delta_{12}$ and $aL/2E$, we find
\begin{eqnarray}
\label{eq:AB-mu}
&\left\{
\begin{array}{lll}
A^{\mu} &=& - \displaystyle\frac{aL}{\Delta_{13}E} 
\displaystyle\frac{1 - 2\sin^2 \theta_{\rm atm}}{\cos^2 \theta_{\rm atm}}
\sin^2 \theta_{\rm rct}\,,\\
B^{\mu} &=& \displaystyle\frac{aL}{4E} \displaystyle\frac{1-2\sin^2 \theta_{\rm atm}}{\cos^2 \theta_{\rm atm}}
\sin^2 \theta_{\rm rct}
\\
 &&- \displaystyle\frac{\Delta_{12}}{2}
\left(  \cos^2 \theta_{\rm sol} 
       + \tan^2 \theta_{\rm atm}\sin^2 \theta_{\rm sol}\sin^2\theta_{\rm rct}
       - \tan \theta_{\rm atm}\sin 2\theta_{\rm sol}\sin \theta_{\rm rct}\cos\dmns
\right)\,,
\end{array}
\right.\\
\label{eq:AB-e}
&\!\!\!\!\!\!\!\!\!\!\!\!\!\!\!\!\!\!\!\!\!\!\!\!\!\!\!\!\!\!\!\!\!\!\!\!
\left\{
\begin{array}{l}
A^e = \displaystyle\frac{aL}{\Delta_{13}E} (1-2\sin^2 \theta_{\rm rct}) 
        -\displaystyle\frac{\Delta_{12}}{2} \displaystyle\frac{\sin2\theta_{\rm sol}}{\tan \theta_{\rm atm} \sin\theta_{\rm rct}}\sin \dmns \,,\\
B^e = - \displaystyle\frac{aL}{4E}(1-2\sin^2 \theta_{\rm rct})
         + \displaystyle\frac{\Delta_{12}}{2}
                       \left(\displaystyle\frac{\sin2\theta_{\rm sol}}{2\tan \theta_{\rm atm} \sin\theta_{\rm rct}}\cos \dmns - \sin^2 \theta_{\rm sol} \right) \,.
                       \end{array}
\right.\end{eqnarray}
The angles
are expressed as the terms of the $3\times3$ MNS matrix \cite{mns} elements
$\sin^2 \theta_{\rm rct} = |U_{e3}|^2$,
$\sin^2 \theta_{\rm atm} = |U_{\mu3}|^2$,
$\sin^2 2\theta_{\rm sol} = 4|U_{e1}U_{e2}|^2$
as in refs.\cite{hagiwara-okamura, t2b}.
$\Delta_{13} > 0$ for the normal hierarchy, $\Delta_{13} < 0$ for the inverted hierarchy,
and $\Delta_{12} \approx  |\Delta_{13}| / 30$ from the constraints eqs.~(\ref{atm-data}) and (\ref{sol-data}).
The term $a$ controles the matter effect \cite{matter effect},
\begin{equation}
a = 2\sqrt{2} G_F E n_e = 7.56 \times 10^{-5} {\rm eV}^2
\left( \frac{\rho}{{\rm g/cm}^3}\right)
\left( \frac{E}{{\rm GeV}} \right) \,,
\label{eq:matter-effect}
\end{equation}
where $n_{e}$ is the number density of the electron and $\rho$ is the matter density.
The magnitude of $aL/2E$ is about 0.17 at SK, while it is about 0.57 at Korea with $\rho=3.0 {\rm g/cm}^3$.
By inserting the typical values of the observed parameters eqs~(\ref{atm-data}) and (\ref{sol-data}),
we find
\begin{eqnarray}
&\left\{
\begin{array}{lll}
A^{\mu} &\sim& 0 \,,\\
B^{\mu} &\sim& -
\left[0.037 + 0.0004 \left(\displaystyle\frac{\sin^2 2\theta_{\rm rct}}{0.10}\right) - 0.008 
\left(
\displaystyle\frac{\sin^2 2\theta_{\rm rct}}{0.10}
\right)^{1/2}
\cos \dmns 
\right]
\displaystyle\frac{|\Delta_{13}|}{\pi} \,,
\end{array}
\right.
\label{eq:AB-mu+}
\\
&\!\!\!\!\!\!\!\!\!\!\!\!\!\!\!\!\left\{
\begin{array}{lll}
A^{e} &\sim& 0.11
\displaystyle\frac{\pi}{\Delta_{13}}
\displaystyle\frac{L}{295{\rm km}}
-\left[0.49
\left( \displaystyle\frac{0.10}{\sin^2 2\theta_{\rm rct}} \right)^{1/2}
\sin \dmns
\right]
\displaystyle\frac{|\Delta_{13}|}{\pi} \,,\\
B^{e} &\sim&
- 0.08
\left( \displaystyle\frac{L}{295{\rm km}} \right)
+\left[0.24 \left( \displaystyle\frac{0.10}{\sin^2 2\theta_{\rm rct}} \right)^{1/2} \cos \dmns - 0.016 \right] 
\displaystyle\frac{|\Delta_{13}|}{\pi} \,,
\end{array}
\right.
\label{eq:AB-e+}
\end{eqnarray}
near the oscillation maximum, $|\Delta_{13}| \sim \pi$.
Because the magnitude of $A^{\mu}$ and $B^{\mu}$ are very small,
the $\nu_{\mu}$ survival rate is rather insensitive to the matter effect and subleading terms of order $\Delta_{12}$.
Accordingly, measurements of the $\nu_{\mu}$ survival probability improve the constraints on $|m^2_3 - m^2_1|$ and $\sin^2 2 \theta_{\rm atm}$.
On the other hand,
both
$A^e$ and $B^e$ can affect the $\nu_{\mu} \rightarrow \nu_e$ oscillation probability significantly.
Most importantly, the magnitude of the transition probability receives the correction term from the matter effect whose sign follows that of $m^2_3 - m^2_1$ and whose magnitude grows with $L$ near the oscillation maximum, $|\Delta_{13}| \sim \pi$.
If we define the difference of the $\nu_{\mu} \rightarrow \nu_e$
transition probability between at Korea and SK, it can be expressed near the oscillation maximum as
\begin{equation}
\begin{array}{lll}
\Delta P_{\rm normal} &=& P_{\mu \rightarrow e}(L_{\rm far}, \Delta_{13} = + \pi) - P_{\mu \rightarrow e}(L_{\rm near}, \Delta_{13} = + \pi) \,,\\
\Delta P_{\rm inverted} &=& P_{\mu \rightarrow e}(L_{\rm far}, \Delta_{13} = - \pi) - P_{\mu \rightarrow e}(L_{\rm near}, \Delta_{13} = - \pi)\,,
\end{array}
\label{eq:dP}
\end{equation}
respectively, for the normal hierarchy ($\Delta_{13} = \pi$) and the inverted one ($\Delta_{13} = -\pi$).
The difference is positive, and can be expressed as
\begin{eqnarray}
\Delta P_{\rm normal} - \Delta P_{\rm inverted} &\sim&
8 \sin^2 \theta_{\rm atm} \sin^2 \theta_{\rm rct}
\left( \displaystyle\frac{a L_{\rm far}}{\pi E_{\rm far}} 
- \displaystyle\frac{a L_{\rm near}}{\pi E_{\rm near}} \right) \nonumber \\
&\sim& 0.01
\left( \displaystyle\frac{\sin^2 2\theta_{\rm rct}}{0.10} \right)
\left( \displaystyle\frac{L_{\rm far} - L_{\rm near}}{295 {\rm km}}\right)
\,.
\label{eq:difference-dP}
\end{eqnarray}
The difference grows linearly with the distance, $L_{\rm far}$, as long as the oscillation maximum is covered by the flux.
The ability of excluding the inverted hierarchy ($\Delta_{13} = - \pi$) is then determined by the error of the $\Delta P_{\rm normal}$, which can be estimated as
\begin{eqnarray}
\delta (\Delta P) &=&
\left[
\left(
\delta P_{\mu \rightarrow e}(L_{\rm near})
\right)^2
+
\left(
\delta P_{\mu \rightarrow e}(L_{\rm far})
\right)^2
\right]^{1/2}
\nonumber \\
&=&
\left[
\left(\displaystyle\frac{P_{\mu \rightarrow e}(L_{\rm near})}{\sqrt{N_e^{\rm near}}}\right)^2
+
\left(
\displaystyle\frac{P_{\mu \rightarrow e}(L_{\rm far})}{\sqrt{N_e^{\rm far}}}
\right)^2
\right]^{1/2}
\,.
\label{eq:deltaP}
\end{eqnarray}
Here $N_e$ is the number of $\nu_e$ appearance events.
$N_e^{\rm far}/N_e^{\rm near}$ can be expressed as
\begin{equation}
\displaystyle\frac{N_e^{\rm far}}{N_e^{\rm near}}=
\displaystyle\frac{V_{\rm far}}{V_{\rm near}} 
\,
\displaystyle\frac{
\Phi_{\rm far}(E_{\nu}~{\rm a}t~\Delta_{13} = \pi, L = L_{\rm far})
}
{
\Phi_{\rm near}(E_{\nu}~{\rm at}~\Delta_{13} = \pi , L = L_{\rm near})
} \,,
\end{equation}
where $V$ denotes the fiducial volume of the detector
and $\Phi(E_{\nu},~L)$ is the neutrino beam flux at $L$,
which is proportional to $(1/L)^2$.
The neutrino cross section of Quasi-Elastic events is almost 
the constant in the 0.7 - 10 GeV region.
Typical event number, $N_e^{\rm near}$,
for $\sin^2 2\theta_{\rm rct} = 0.1$ and $\dmns = 0^{\circ}$ during the 5 years running at SK 
is about 200;
see a few bins around 0.5 GeV in Fig.~(\ref{fig:event-numbers}).
We therefore estimate
significance of excluding the fake hierarchy as
\begin{equation}
\displaystyle\frac{\Delta P_{\rm normal} - \Delta P_{\rm inverted}}{\delta ( \Delta P )}
= 2.8
\left( 
          \frac{\sin^2 2\theta_{\rm rct}}{0.10} 
          \right)^{1/2}
\left( \displaystyle\frac{L_{\rm far} - L_{\rm near}}{295{\rm km}}\right)
\left[
       1+0.225\left(\frac{L_{\rm far}}{295{\rm km}}\right)^{2}
       \frac{100 {\rm kt}}{V_{\rm far}}
\right]^{-1/2}
\,.
\end{equation}
We find that when we put a 100~kt detector at $L=1000$~km,
the significance can exceed 3.5 in this rough estimate.
\par
As of March 2005,
there is no proposal to construct a huge neutrino detector in Korea.
In our case study,
we consider a 100~kt level detector,
in order to compensate for the decrease in the neutrino flux 
which is about $(300~{\rm km}/1,000~{\rm km})^2 \sim 1/10$ of that at SK.
We adopt a Water \cerenkov detector because it allows us to distinguish 
clearly the $e^{\pm}$ events
from $\mu^{\pm}$ events.
We use the Charged-Current-Quasi-Elastic (CCQE) events in our analysis,
because they allow us to reconstruct the neutrino energy
by measuring the strength and the orientation of the \cerenkov lights.
Since the Fermi-motion effect of the target dominates 
the uncertainty of the neutrino energy reconstruction,
which is about 80 MeV,
in the following analysis we take the width of the energy bin as 
$\delta E_\nu=200$~MeV for $E_\nu > 400$ MeV. 
The signals in the $i$-th energy bin,
$E_\nu^i = 200{\rm MeV} \times (i+1) < E_{\nu} < E_{\nu}^i + \delta E_{\nu}$,
are then calculated as
\begin{equation}
\label{eq:N}
N_\alpha^{i} (\nu_\mu)=
 M N_A
 \int_{E_\nu^i}^{E_\nu^{i}+\delta E_\nu}
 \Phi_{\nu_\mu}(E)~
 P_{\nu_\mu\to\nu_\alpha}(E)~ 
 \sigma_\alpha^{QE}(E)~
 dE\,,
\end{equation}
where
$P_{\nu_{\mu}\rightarrow \nu_{\alpha}}$
is the neutrino oscillation probability including the matter effect, 
$M$ is the mass of detector,
$N_{A} = 6.017\times10^{23}$ is Avogadro constant,
$\Phi_{\nu_\mu}$ is the $\nu_{\mu}$ flux from J-PARC,
and
$\sigma_\alpha^{QE}$ is the CCQE cross section per nucleon in water.
In this study, we assume that the fiducial volume of Super-Kamiokande 
is 22.5~kt, and that of a detector in Korea is 100~kt.
We also assume that the detection efficiencies of both detectors for the CCQE events is
$100\%$.
The typical event numbers with $\sin^2 2\theta_{\rm rct} = 0.1$ and $\dmns = 0^{\circ}$ is shown by Fig.~{\ref{fig:event-numbers}}. 
\par
We consider the following background events 
for the signal $e$- and $\mu$-like events
\begin{eqnarray}
\label{eq:BG_e}
N_{\alpha}^{i,{\rm BG}}
&=&
N_{\alpha}^{i}(\nu_e) +
N_{\bar{\alpha}}^{i}(\bar{\nu}_e) +
N_{\bar{\alpha}}^{i}(\bar{\nu}_\mu)\,,
{\mbox{\hspace*{5ex}}}
(\alpha = e\,, \mu)\,.
\end{eqnarray}
The three terms represent the contribution from the secondary neutrino 
flux of the  $\nu_\mu$ primary beam,
which are calculated as in eq.~(\ref{eq:N})
where $\Phi_{\nu_{\mu}}(E)~$ is replaced by $\Phi_{\nu_{\beta}}(E)$ for
$\nu_{\beta} = \nu_e\,,\bar{\nu}_e\,,\bar{\nu}_{\mu}$.
After summing up these background events,
the $e$-like and $\mu$-like events for the $i$-th bin are obtained as
\begin{equation}
\label{signal}
 N_{\alpha}^{i} = N_{\alpha}^{i}(\nu_\mu) + N_{\alpha}^{i,{\rm BG}}
\,,
{\mbox{\hspace*{5ex}}}
(\alpha = e\,, \mu)\,.
\end{equation}
\par
Since our concern is the possibility to distinguish the neutrino mass 
hierarchy and to measure \sir and the CP phase,
we study how the above `data' can constrain the model parameters by using the $\chi^2$ function
\begin{equation}
\label{chi^2 define}
\Delta\chi^2 = \chi^2_{\rm SK} + \chi^2_{\rm Kr} + \chi^2_{\rm sys} 
+ \chi^2_{\rm para}\,.
\end{equation}
Here the first two terms, $\chi^2_{\rm SK}$ and $\chi^2_{\rm Kr}$,
measure the parameter dependence of the fit to the SK and the Korean 
detector data,
\begin{eqnarray}
\label{eq:chi^2event}
 \chi^2_{\rm SK,Kr}
 = \sum_{i} \left\{
\left(
\displaystyle\frac
{(N_e^{i})^{\rm fit} - N_e^{i}}
{\sqrt{N^i_e}}
\right)^2
+
\left(
\displaystyle\frac
{(N_\mu^{i})^{\rm fit} - N_\mu^{i}}
{\sqrt{N^i_{\mu}}}
\right)^2
\right\}\,,
\end{eqnarray}
where the summation is over all bins
from 0.4 GeV up to 5.0 GeV for $N_{\mu}$,
1.2 GeV for $N_{e}~$at SK,
and 2.8GeV for $N_{e}~$at Korea.
These upper bounds are chosen such that most of the bins used in our
analysis contain more than 10 events.
Here $N^i_{\mu,e}$ is the calculated number of events
in the $i$-th bin,
and its square root gives the statistical error.
In our analysis,
we calculate $N^i_{\mu,e}$ 
by assuming the
following input (`true') parameters:
\begin{equation}
\left.
\begin{array}{l}
(m^2_3 - m^2_1)^{\rm true} = 2.5 \times 10^{-3}~\mbox{eV}^2 ~~ ( >0) \,,\\
(m^2_2 - m^2_1)^{\rm true} = 8.3 \times 10^{-5}~\mbox{eV}^2\,, \\
\sin^2 \theta_{\rm atm}^{\rm true} = 0.5 \,,  \\
\sin^2 2\theta_{\rm sol}^{\rm true} = 0.84 \,, \\ 
\sin^2 2\theta_{\rm rct}^{\rm true} = 0.1 \,, ~~0.06 \,, \\
\dmns^{\rm true} = 0^{\circ},~ 90^{\circ},~ 180^{\circ},~ 270^{\circ} \,,
\end{array}
\right\}
\label{eq:input}
\end{equation}
with the constant matter density, $\rho^{\rm true}=2.8~{\rm g/cm}^3$ for T2K
and $\rho^{\rm true}=3.0~{\rm g/cm}^3$ for the Tokai-to-Korea experiments.
Note that in eq.~(\ref{eq:input}),
we assume the normal hierarchy ($m^2_3 - m^2_1 >0$)
and
examine several input values of \sir and \cpdelta.
\par
$N_{i}^{\rm fit}$ is calculated by allowing the model parameters
to vary freely
and by allowing for systematic errors.
In our analysis, we consider 4 types of systematic errors.
The first ones are for the overall normalization of each neutrino flux,
for which we assign $3\%$ errors,
\begin{equation}
f_{\nu_\beta} = 1 \pm 0.03\,,
\end{equation}
for $\nu_{\beta} = \nu_{e},~\bar{\nu}_{e},\nu_{\mu},~\bar{\nu}_{\mu}$,
which are taken common for T2K and the Tokai-to-Korea experiments.
The second systematic error is for the uncertainty in the matter density,
for which we allow $3\%$ overall uncertainty along the baseline, 
independently for T2K ($f^{\rm SK}_{\rho}$) and
the Tokai-to-Korea experiment ($f^{\rm Kr}_{\rho}$):
\begin{equation}
\rho_{i}^{\rm fit} = f^{i}_{\rho}\,\rho^{\rm true}_i
\hspace{5ex}
(i = {\rm SK,~Kr}) \,.
\end{equation}
The third uncertainty is for the CCQE cross section,  
\begin{equation}
\sigma^{\rm QE,~fit}_{\alpha} = f^{\rm QE}_{\alpha}\, 
\sigma^{\rm QE,~true}_{\alpha} \,.
\end{equation}
Since $\nu_e$ and $\nu_\mu$ QE cross sections are expected to be very 
similar theoretically, we assign a common overall error of $3\%$ for 
$\nu_e$ and $\nu_{\mu}$ 
($f_e^{\rm QE} = f_\mu^{\rm QE} \equiv f_\ell^{\rm QE}$), 
and an independent $3\%$ error for $\bar{\nu}_e$ and $\bar{\nu}_\mu$ QE 
cross sections 
($f_{\bar{e}}^{\rm QE} = f_{\bar{\mu}}^{\rm QE} 
\equiv f_{\bar{\ell}}^{\rm QE}$).  
The last one is the uncertainty of the fiducial volume,
for which we assign $3\%$ error independently for T2K 
($f_{\rm V}^{\rm SK}$) and the
Tokai-to-Korea experiment ($f_{\rm V}^{\rm Kr}$).
$N_{\alpha}^{i,{\rm fit}}$ is then calculated as
\begin{eqnarray}
N_\alpha^{i,{\rm fit}}(\nu_\beta) &=&
f_{\nu_\beta}\, f_{\alpha}^{\rm QE}\, f_{\rm V}^{\rm SK,Kr}\,
N_\alpha^{i}(\nu_\beta)\,, 
\end{eqnarray}
and $\chi^2_{\rm sys}$ has four terms;
\begin{equation}
\label{chisq-sys}
 \chi^2_{\rm sys} = 
\sum_{\alpha = e,\bar{e},\mu,\bar{\mu}}
\left(
\displaystyle\frac{f_{\nu_{\alpha}}-1}{0.03}
\right)^2
+
\sum_{\alpha = l, \bar{l}}
\left(
\displaystyle\frac{f^{\rm QE}_{\alpha}-1}{0.03}
\right)^2
+
\sum_{ i = {\rm SK,~Kr}}
\left\{
\left(
\displaystyle\frac{f^{i}_{\rho}-1}{0.03}
\right)^2
+
\left(
\displaystyle\frac{f^{i}_{\rm V}-1}{0.03}
\right)^2
\right\}\,.
\end{equation}
In short,
we assign $3\%$ errors for the normalization of each neutrino flux,
the $\nu_e$ and $\bar{\nu}_e$ CCQE cross sections,
the effective matter density along the base line,
and for the fiducial volume of SK and the Korean detector.
Finally, $\chi^2_{\rm para}$ accounts for the present constraints 
on the model parameters, summarized in eqs.~(1) and (2):
\begin{eqnarray}
\label{chisq-para}
\chi^2_{\rm para}
&=&
\left(
\displaystyle\frac{ |(m_3^2 - m_1^2)^{\rm fit}| - |(m_3^2 - m_1^2)^{\rm true}|}
{0.5 \times 10^{-3}}
\right)^2
+
\left(
\displaystyle\frac{(m_2^2 - m_1^2)^{\rm fit} - (m_2^2 - m_1^2)^{\rm true}}
{ 0.6 \times 10^{-5}}
\right)^2\nn\\
&&+
\left(
\displaystyle\frac
{\sin^2 2\theta_{\rm atm}^{\rm fit}- \sin^2 2\theta_{\rm atm}^{\rm true}}
{0.04}
\right)^2
+
\left(
\displaystyle\frac
{\sin^22\theta_{\rm sol}^{\rm fit}- \sin^22\theta_{\rm sol}^{\rm true}}
{0.07}
\right)^2\,.
\end{eqnarray}
Here we interpret the $90\%$ CL lower bound on \sia in eq.~(\ref{atm-data})
as the $1.96\sigma$ constraint from $\sin^2 2\theta_{\rm atm}$ is greater than 0.96,
and
the asymmetric error for $\tan^2 \theta_{\rm sol}$ in eq.~(\ref{sol-data}) 
has been made more
symmetric for \sis.
We do not include the bounds on \sir in eq.~(\ref{rct-data}) in our 
$\Delta\chi^2$ function.
In total, our $\Delta\chi^2$ function depends on 16 parameters,
the 6 model parameters and the 10 normalization factors.
\par
\begin{figure}
\begin{center}
\includegraphics[angle = 0 ,width=7.5cm]{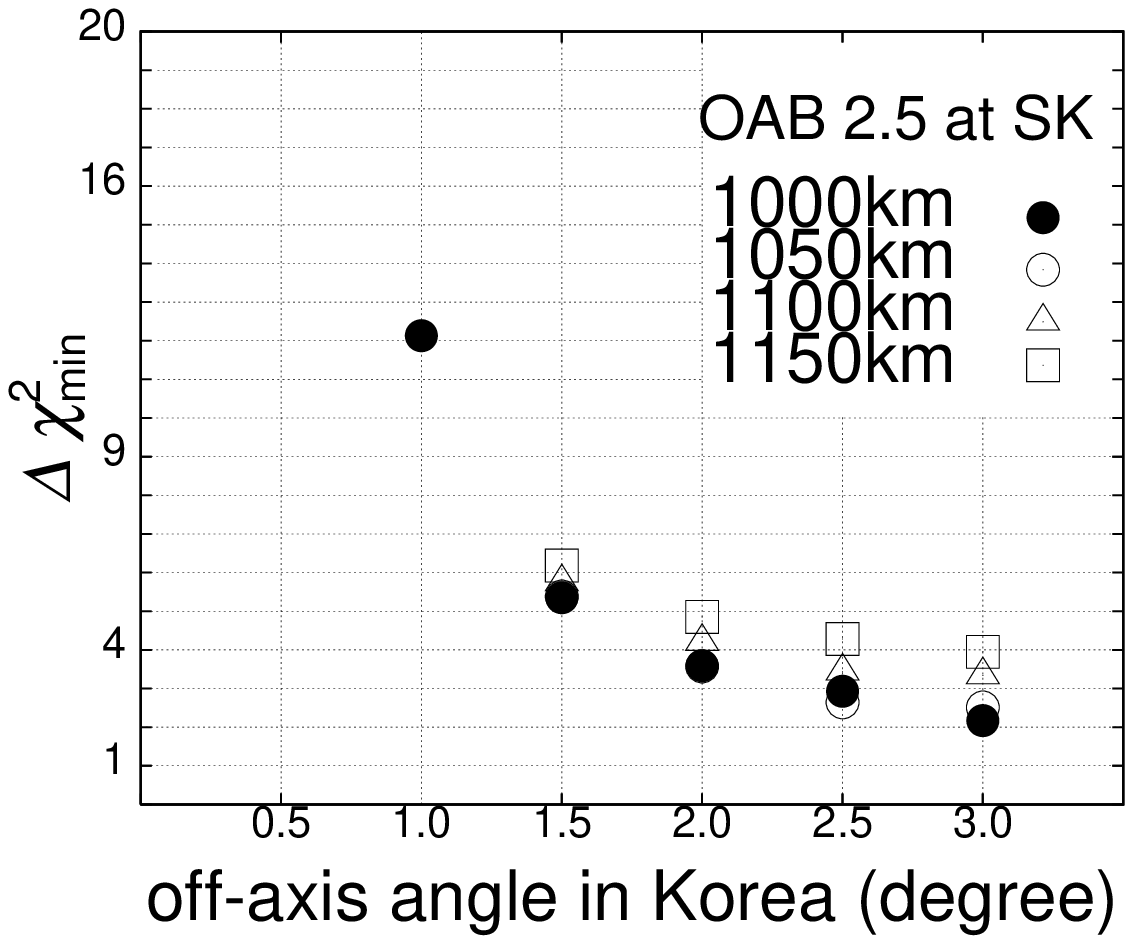}
\includegraphics[angle = 0 ,width=7.5cm]{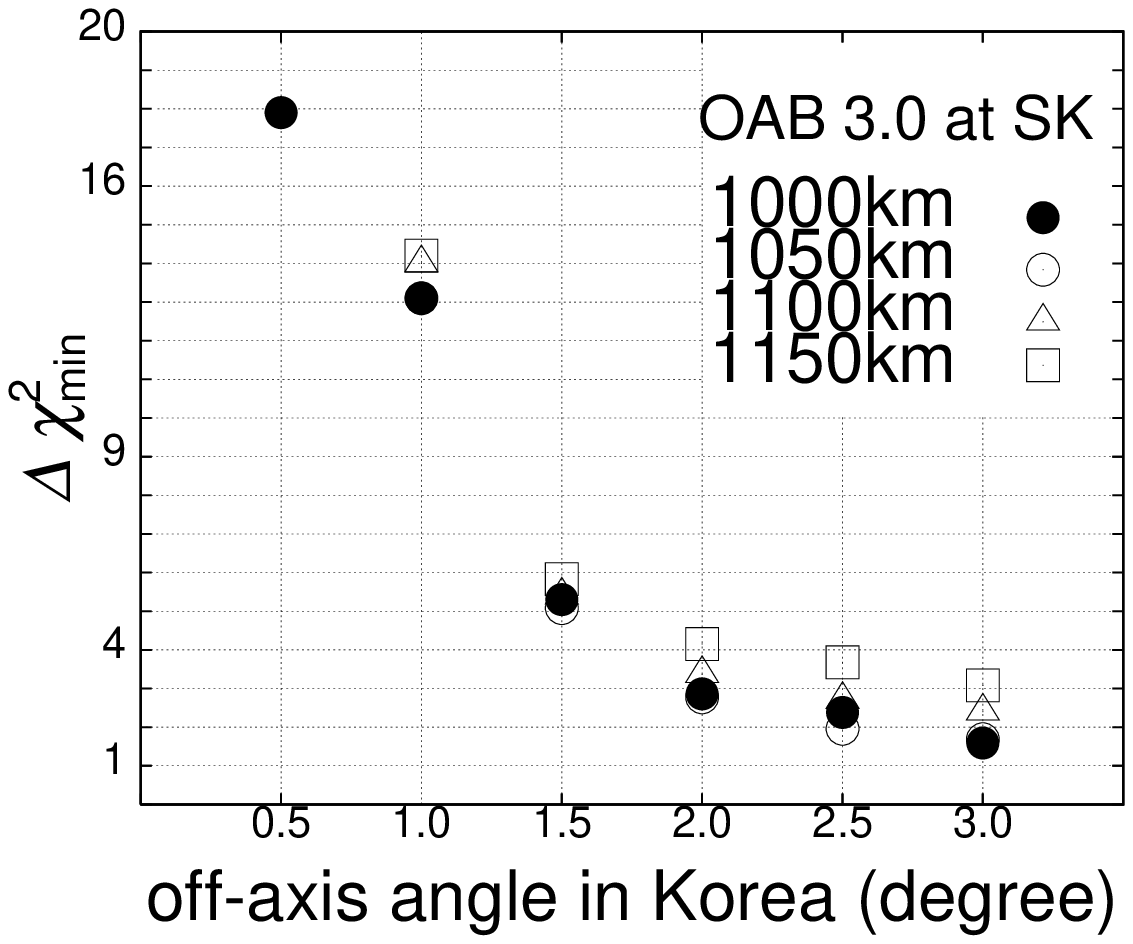}
\end{center}
\caption{Minimum $\Delta\chi^2$ as functions of the off-axis angle and
the base-line length from J-PARC at Tokai,
when the normal hierarchy 
($m_3^2 - m_1^2 = 2.5 \times 10^{-3} {\rm eV}^2 > 0$)
is assumed in generating the events,
and the inverted hierarchy
($m_3^2 - m_1^2 < 0$) is assumed in the fit. 
The left-hand figure is for the $2.5^\circ$ off-axis beam at SK,
and the right-hand one is for the $3.0^\circ$ beam.
}
\label{fig:hierarchy-L-angle}
\end{figure}
First, 
we search for the best place for the detector in Korea and
the best combination of the off-axis angle for SK and the Korean
detector to determine the sign of $m_{3}^2 - m_{1}^2$.
We show in Fig.\ref{fig:hierarchy-L-angle} the minimum $\Delta\chi^2$ as 
functions of the off-axis angle
and the base-line length in Korea,
when the data, $N^{i}_{\alpha}$,
are generated for the normal hierarchy,
$m_{3}^{2} - m_{1}^{2} = 2.5 \times 10^{-3} {\rm eV}^2 > 0$, 
eq.~(\ref{eq:input}),
and the fit has been performed by assuming the inverted hierarchy,
$m_{3}^{2} - m_{1}^{2} < 0$.
We set $\sin^2 2\theta_{\rm rct}^{\rm true} = 0.10 $ and 
$\dmns^{\rm true} = 0^{\circ}$ in this analysis.
The left hand figure shows the minimum $\Delta\chi^2$ for the $2.5^{\circ}$ 
off-axis beam at SK,
and the right hand one is for the $3.0^{\circ}$ off-axis beam at SK.
The four symbols, solid circle, open circle, triangle, and square are 
for $ L = 1000$km, 1050km, 1100km, and 1150km, respectively.
When the off-axis angle at SK is $2.5^{\circ}$,
the $0.5^{\circ}$ beam does not reach the Korean coast; see Fig.~\ref{fig:map}.
It is clear from these figures that the best discriminating power is obtained
for the combination $L$ = 1000km and $0.5^{\circ}$, 
which is available only when the off-axis angle at SK is $3.0^{\circ}$ 
(right figure).  
For this combination, we can distinguish the inverted hierarchy from 
the normal one at more than 4$\sigma$ level.
These figures show that longer base-line gives larger
minimum $\Delta\chi^2$ for the same off-axis angle.
This is because of the increase in the matter effect.
For the same base-line length,
lower off axis angle beams give better discriminating power.
This is because the neutrino flux with smaller off-axis angle is harder 
\cite{t2k, ichikawa}, and the stronger matter effect to help us to
distinguish the neutrino mass hierarchy \cite{lipari99,barger01,t2b}.
\par
Here after,
we study the prospect for measuring the sign of $m_3^2 - m_1^2$
in more detail for the best combination, 
$L=1000$km and $0.5^\circ$ in Korea, and $3.0^\circ$ for SK.
\begin{figure}
\begin{center}
\includegraphics[angle = 0 , width=7.5cm]{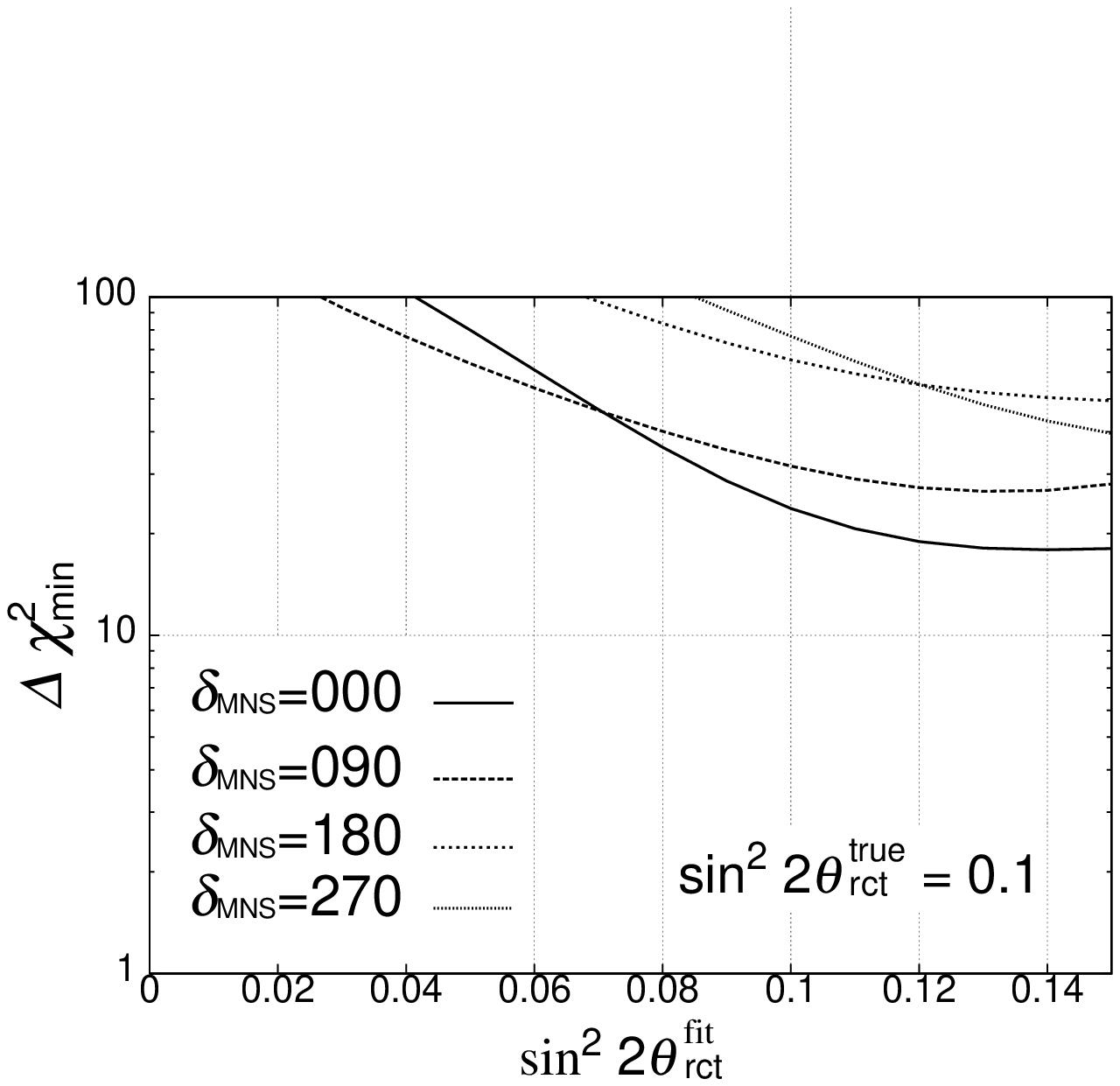}
\includegraphics[angle = 0 , width=7.5cm]{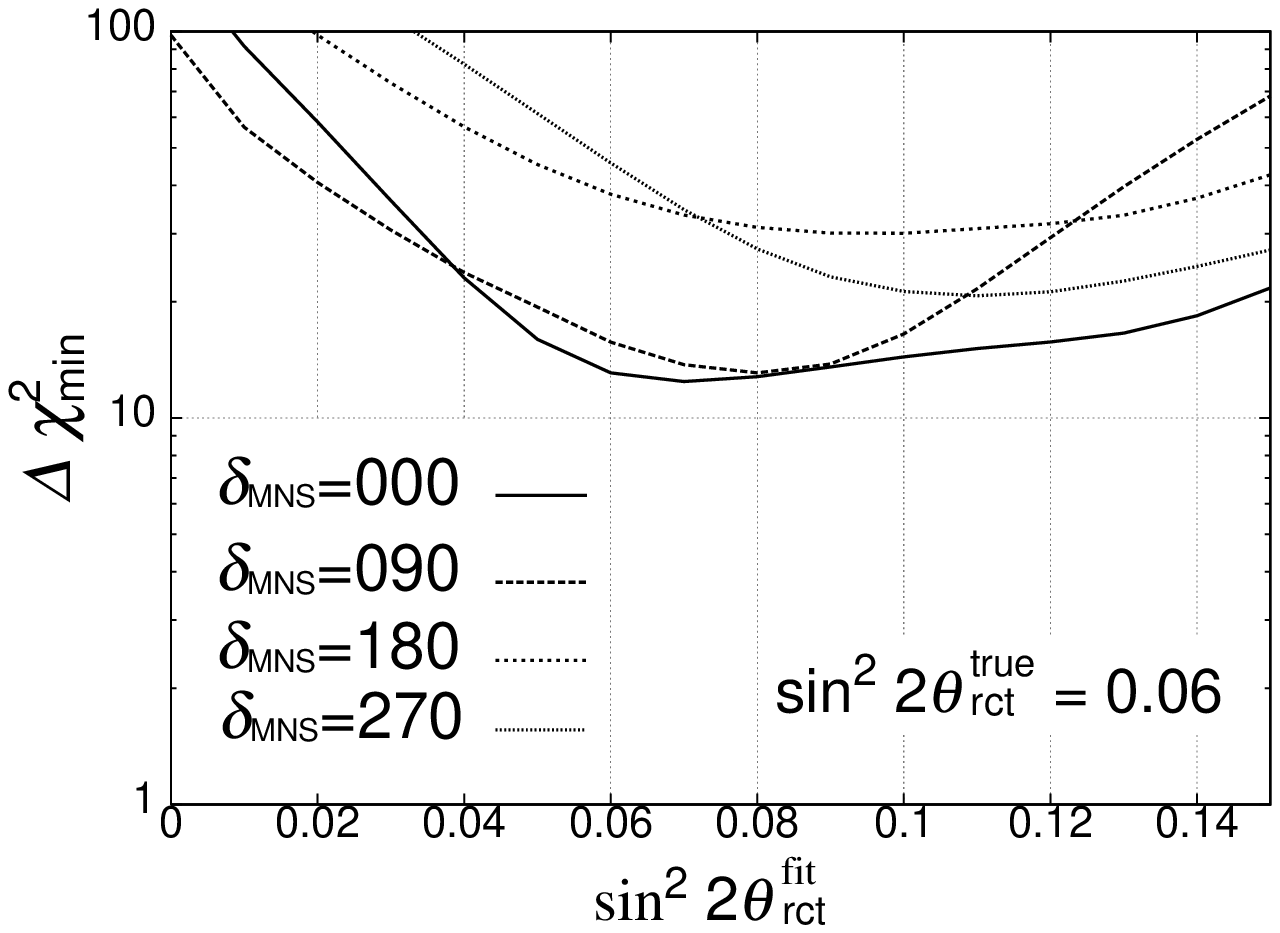}
\end{center}
\caption{Minimum $\Delta\chi^2$ as a function of $\sin^22\theta_{\rm rct}^{\rm fit}$
when the normal hierarchy 
($m_3^2 - m_1^2 = 2.5 \times 10^{-3} ~{\mbox eV}^2 > 0$)
is assumed in generating the events,
and the inverted hierarchy
($m_3^2 - m_1^2 < 0$) is assumed in the fit. 
The 4 lines are for 4 input CP phase values,
$\dmns^{\rm true} = 0.0^{\circ}$ (solid),
$90^\circ$ (long-dashed),
$180^\circ$ (short-dashed) and
$270^\circ$ (dotted).
The left figure is for $\sin^22\theta_{\rm rct}^{\rm true}=0.10$
and the right one is for $\sin^22\theta_{\rm rct}^{\rm true}=0.06$.
}
\label{fig:hierarchy-sir-cp}
\end{figure}
Fig.~\ref{fig:hierarchy-sir-cp} shows the minimum $\Delta\chi^2$ as functions 
of the fitting parameter
$\sin^22\theta_{\rm rct}^{\rm fit}$ by assuming the inverted hierarchy,
when the normal hierarchy 
($m_3^2 - m_1^2 = 2.5 \times 10^{-3}~{\mbox eV}^2 > 0$)
is assumed in calculating the event numbers.
There are 4 lines in each figure,
which correspond to 4 input values of the CP phase,
$\dmns^{\rm true} = 0^{\circ}$ (solid),
$90^\circ$ (long-dashed),
$180^\circ$ (short-dashed) and
$270^\circ$ (dotted),
respectively.
The left figure is for $\sin^22\theta_{\rm rct}^{\rm true}=0.1$
and the right one is for $\sin^22\theta_{\rm rct}^{\rm true}=0.06$.
$\Delta \chi^2$ is mainly controled by
the difference of $N^i_e$ between the normal hierarchy and the inverted hierarchy which is proportional to 
$\sin^2 \theta_{\rm rct}$; see \eqref{eq:difference-dP}.
Because $P_{\nu_{\mu} \rightarrow \nu_e}$ in the inverted hierarchy
is smaller than that in the normal hierarchy due to the matter effect,  
the fitting parameter, $\sin^2 2\theta_{\rm rct}^{\rm fit}$, 
tends to be larger than the input value, 
$\sin^2 2\theta_{\rm rct}^{\rm true}$. 
Since large \sir will be constrained more strongly
in the future reactor experiments \cite{double-chooz, kaska},
we can conclude that
the neutrino mass hierarchy will be determined
at even higher confidence level once results from these reactor 
experiments are available.
%
\par
\begin{figure}
\includegraphics[angle = 0 , width = 8cm] {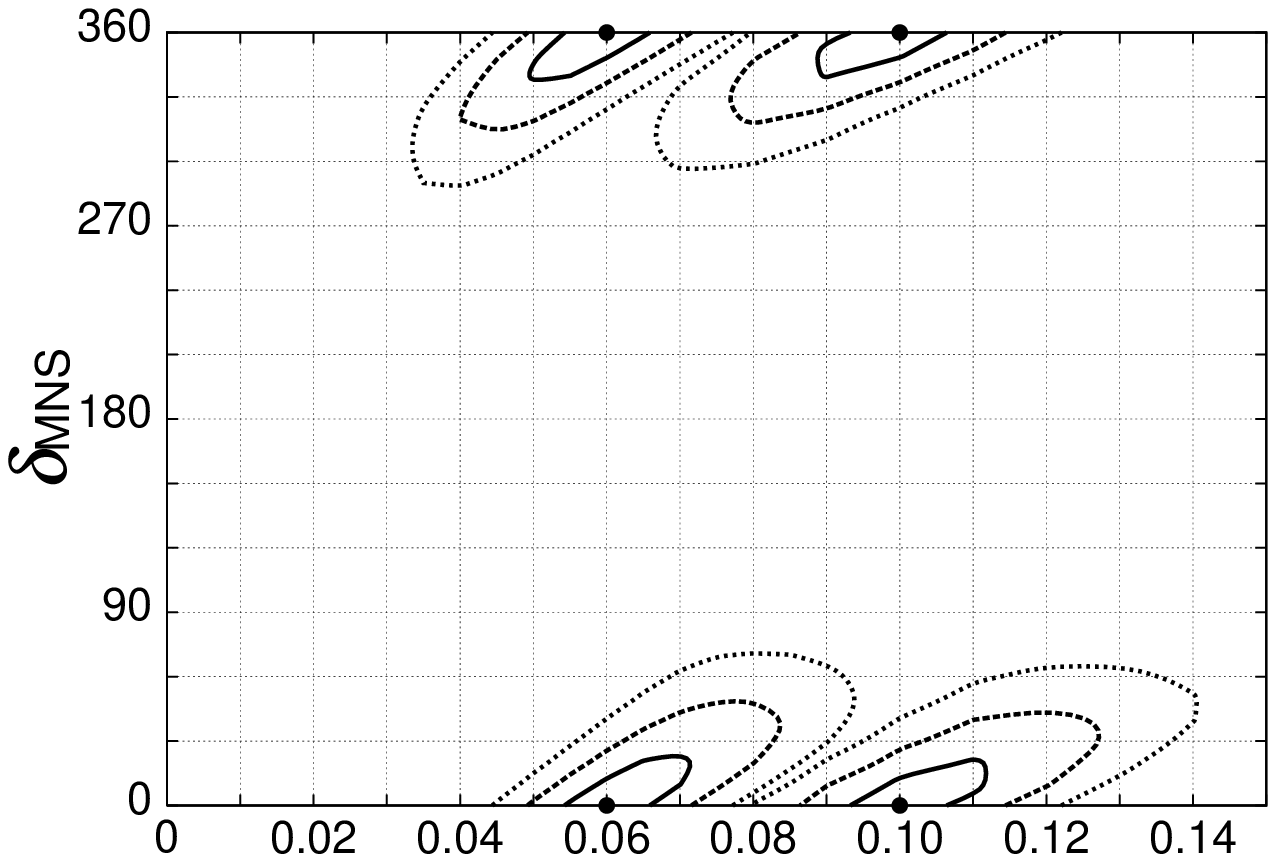}
\includegraphics[angle = 0 , width = 8cm] {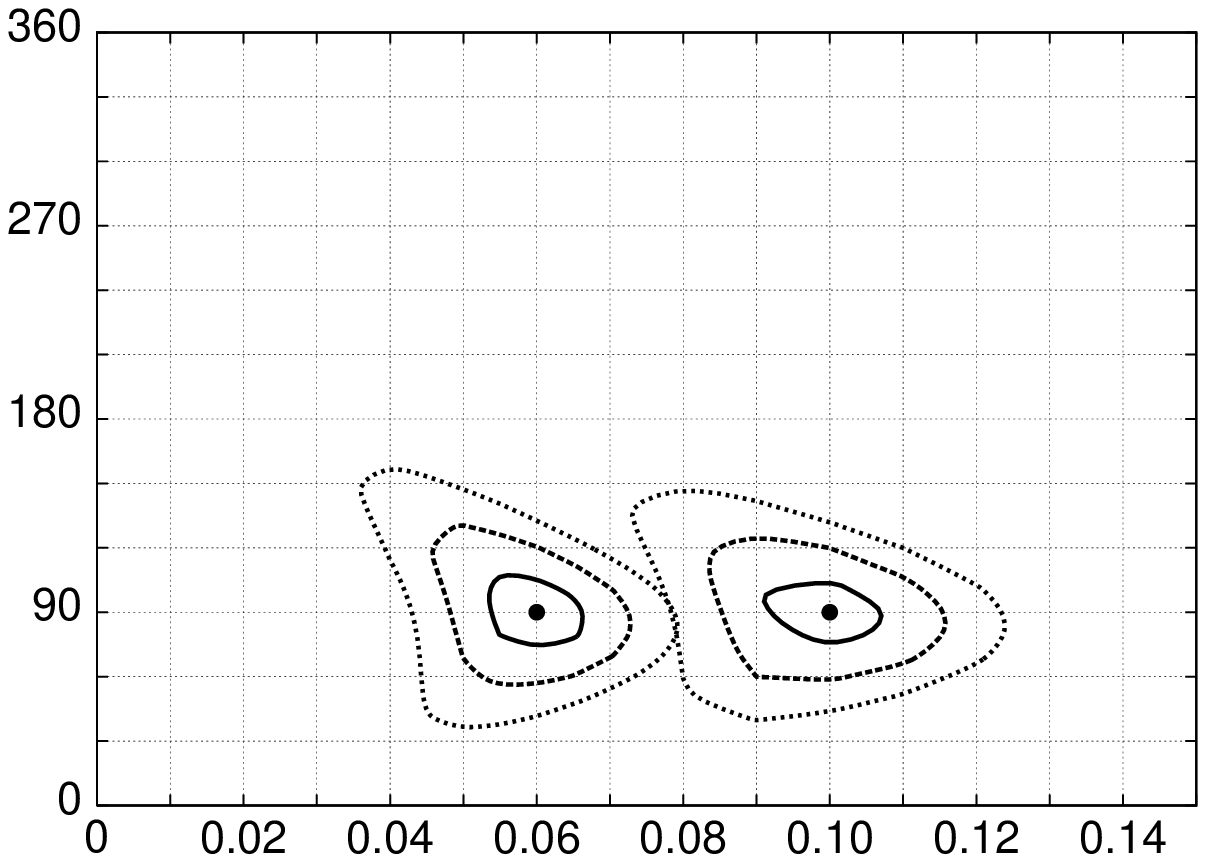}
\includegraphics[angle = 0 , width = 8cm] {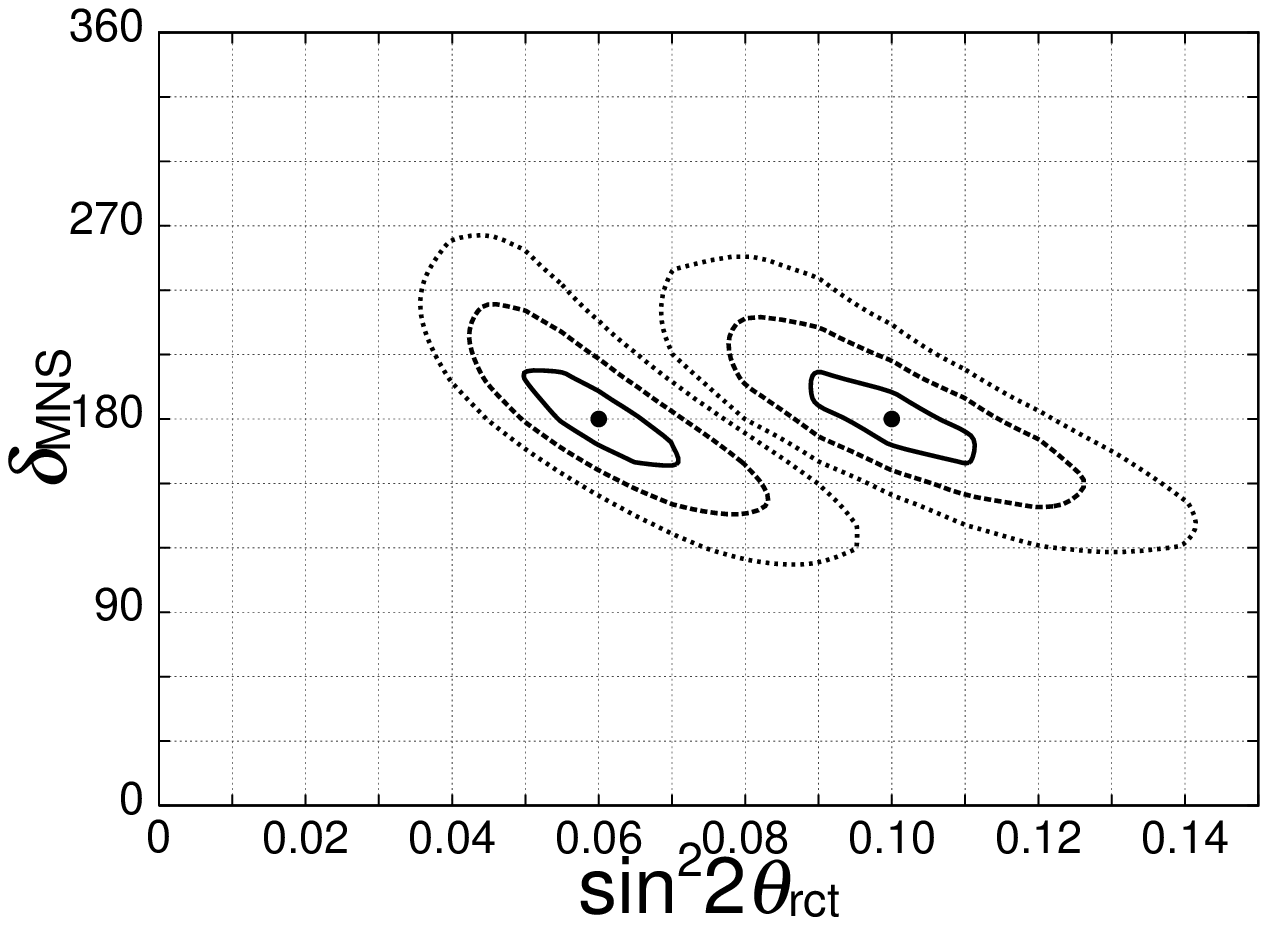}
\includegraphics[angle = 0 , width = 8cm] {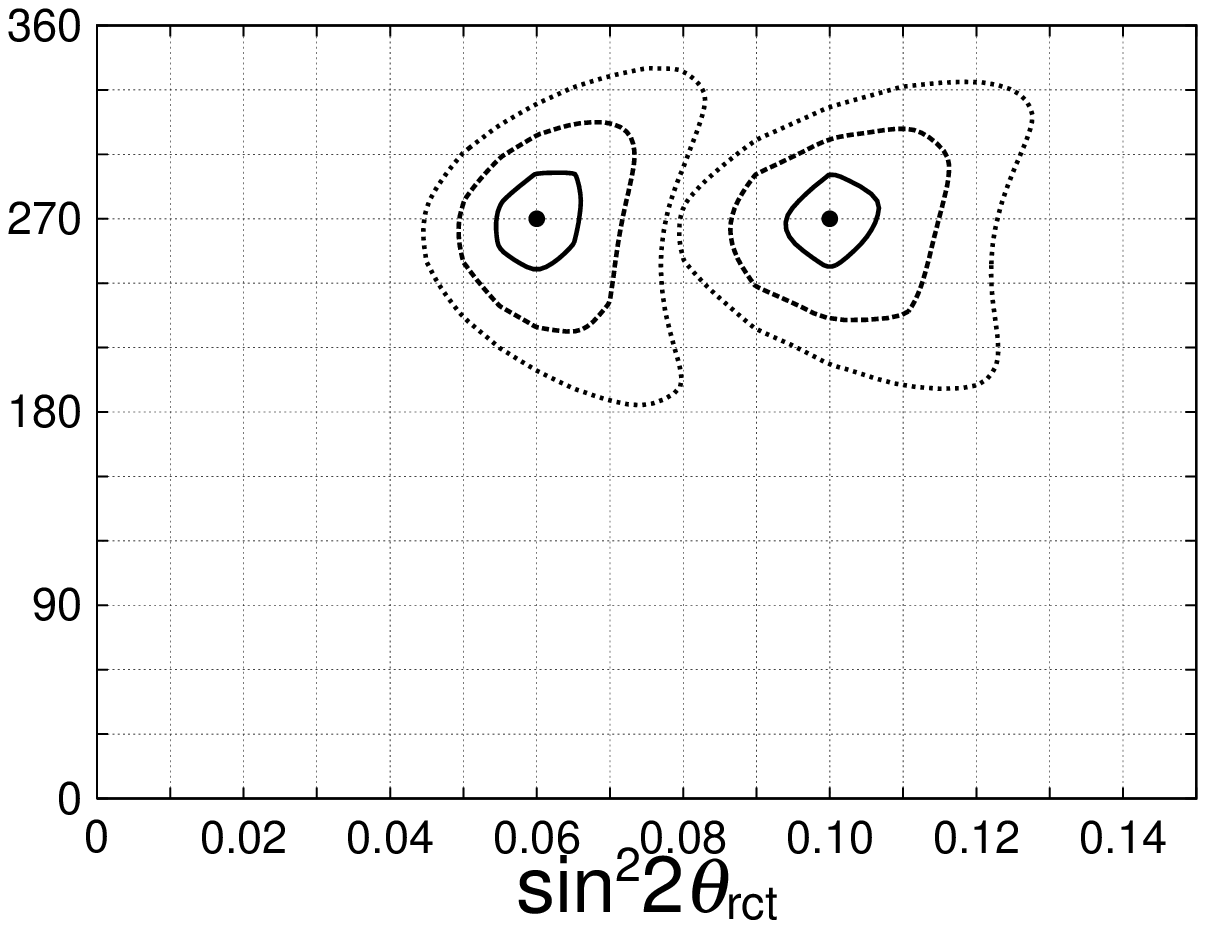}
\caption{
Allowed region in the plane of $\sin^2 2\theta_{\rm rct}^{\rm fit}$ and 
$\dmns^{\rm fit}~$, when the event numbers at SK and Korea are calculated 
for the parameters of eq.~(\ref{eq:input}).
In each figure, the input points 
($\sin^2 2\theta_{\rm rct}^{\rm true}$, $\dmns^{\rm true}$) 
are shown by solid-circles, and the regions where the minimum 
$\Delta\chi^2$ is less than 1, 4, 9 are depicted by solid, dashed and 
dotted boundaries, respectively.
}
\label{fig:cp-contr}
\end{figure}
We also examine the capability of the Tokai-to-Korea LBL experiments
for measuring the CP phase.
We show in Fig.~\ref{fig:cp-contr} regions allowed by this experiment
in the plane of \sir and $\dmns$.
The mean values of the inputs are calculated for the parameters of 
eq.~(\ref{eq:input}).
In each figure, the input points
($\sin^2 2\theta_{\rm rct}^{\rm true}$, $\dmns^{\rm true}$) are shown by
solid-circles for $\sin^2 2\theta_{\rm rct}^{\rm true} =0.10$,
and
$0.06$.
The regions where the minimum $\Delta\chi^2$ is less than 1, 4, 9 are 
depicted by solid, dashed and dotted boundaries, respectively.
Even though we allow the sign of $m_3^2 - m_1^2$ to vary in the fit,
no solution with the inverted hierarchy that satisfy 
$\Delta\chi^2_{\rm min} <9$ appear in the figure.  
\par
From these figures, 
we learn that $\dmns$ can be constrained to $\pm 30^\circ$
at 1$\sigma$ level, when $\sin^22\theta_{\rm rct}^{\rm true}>0.06$.
It is remarkable that we can constrain both $\sin \dmns$ and $\cos \dmns$ without using anti-neutrino experiments.
We can determine $\sin \dmns$ uniquely by measuring the $\nu_{\mu} \rightarrow \nu_e$
oscillation probability near the oscillation maximum both at SK and Korea.
This is because the significant difference in the matter effect term in \eqref{eq:AB-e}
between SK and Korea allows us to resolve
the correlation between $\sin^2 \theta_{\rm rct}$ and $\sin \dmns$
\cite{arafune97,minakata97,barger02}.
As for $\cos \dmns$,
it appears only in the phase shift of the $\nu_{\mu} \rightarrow \nu_e$ oscillation probability;
see the term $B^e$ in eq.~(\ref{eq:AB-e}).
It is therefore important to measure the neutrino energy by CCQE events,
in order to constrain $\cos \dmns$.
%
%
%
%
\par
In this paper,
we study the possibility of solving the degeneracy of
the neutrino mass hierarchy
and
constraining \sir and \cpdelta
by measuring the T2K off-axis beam in Korea.
We find that
by placing a 100~kt level water \cerenkov detector in the east coast of Korea,
we can determine the sign of $m^2_3 - m^2_1$ and 
constrain \sir and $\dmns$ uniquely,
if \sir $\simgt 0.06$ .
\par
Our results are based upon a very simple treatment of the systematic 
errors where $3\%$ overall errors are assigned for all the 10 normalization 
factors of eq.~(\ref{chisq-sys}).  
Even if we enlarge all the systematic errors to $10\%$ except for the matter density uncertainties,
the significance of the mass hierarchy determination is not affected much.
For instance, the $\Delta \chi^2_{\rm min}$ for the combination of the $3^{\circ}$ OAB at SK and the $0.5^{\circ}$
OAB at $L = 1000$ km in \figref{fig:hierarchy-L-angle} is found to reduce from 18 to 16.
Among the potentially serious background which we could not estimate in this paper are;
\begin{itemize}
  \item possible miss-identification of NC $\pi^0$ production as $\nu_e$ CCQE events,
  \item possible miss-identification of soft $\pi$ emission events as $\nu_e$ CCQE events.
\end{itemize}
Although the above uncertainties were found to be rather small at K2K experiments
\cite{NC},
we should expect them to be more serious at high energies.
Dedicated studies of their effects on the neutrino-energy reconstruction efficiency
are mandatory.
In addition,
careful studies including possible energy 
dependence of the flux and cross section uncertainties, location 
dependence of the matter density may be needed to justify the physics case of our 
proposal.
\\
{\it Acknowledgments}\\
We thank our colleagues
Y.~Hayato,
A.K.~Ichikawa,
T.~Ishii,
I.~Kato,
T.~Kobayashi
and
T.~Nakaya,
from whom learn about the K2K and T2K experiments.
We are also grateful to 
Mayumi Aoki,
C.W.~Kim,
Soo-Bong Kim,
and
Yeongduk Kim,
for useful discussions and comments.
The work of KH is supported in part
by the Core University Program of JSPS.

{\it Note added:}\\
When we were finalizing the manuscript for publication, we learned that a 
similar study has been performed by M.~Ishituka \etal~\cite{t2kr-m}.

\end{document}